\documentclass[12pt]{amsart}
\usepackage{amssymb}
\def\mylabel#1{\label{#1}}
\def\myref#1{\ref{#1}}
\def\mycite#1{\cite{#1}}
\def\cop{\Delta}
\def\ot{\otimes}
\def\coo#1{#1_{(1)}}
\def\cot#1{#1_{(2)}}

\def\oh{\frac{1}{2}}
\def\Q{\mathbb Q}
\def\Z{{\mathbb Z}}
\def\C{{\mathbb C}}

\def\CB{{\mathcal B}}
\def\CH{{\mathcal H}}
\def\CF{{\mathcal F}}

\def\CT{{\mathcal T}}

\def\CT{\mathcal T}
\def\CB{\mathcal B}
\def\CC{\mathcal C}
\def\acts{\triangleright}
\DeclareMathSymbol\crossrt{\mathrel}{AMSb}{"6E}
\DeclareMathSymbol\crosslt{\mathrel}{AMSb}{"6F}

\def\id{\hbox{id}}
\def\ts{\otimes}

\newtheorem{lemma}{Proposition}[section]

\newtheorem{corollary}[lemma]{Corollary}
\theoremstyle{definition}
\newtheorem{example}[lemma]{Example}
\newtheorem{definition}[lemma]{Definition}

\newtheorem{observation}[lemma]{Observation}
\newcommand\mapright[1]{\smash{\mathop{\longrightarrow}\limits^{#1}}}
\title{Multiple noncommutative tori and Hopf algebras}
\author{Markus Debert}
\author{Mario Paschke}
\address[M.Debert, M.Paschke]{Institut f\"ur Physik, Johannes-Gutenberg Universit\"at, 55130 Mainz, Germany}
\email[M.Paschke]{paschke@thep.physik.uni-mainz.de}
\email[M.Debert]{debert@zino.physik.uni-mainz.de}
\author{Andrzej Sitarz}
\address[A.Sitarz]{Institute of Physics, Jagiellonian University, Reymonta 4, 30-059 Krak\'ow, Poland}
\curraddr[A.Sitarz]{Laboratoire de Physique Th\'eorique, Batiment 210
Universit\'e Paris XI, 91405 Orsay Cedex, France}
\thanks{${}^\dagger$A.Sitarz acknowledges support by Polish Committee
for Scientific Research, KBN grant 2P03B 023 14 in the early stage
of the work and Marie Curie Fellowship.}
\email[A.Sitarz]{Andrzej.Sitarz@th.u-psud.fr}

\begin{document}
\begin{flushright}
MZ-TH 01-09 \\
LPT-ORSAY 01-28
\end{flushright}
\maketitle
\begin{abstract}
We derive the Kac-Paljutkin finite-dimensional Hopf algebras as
finite fibrations of the quantum double torus and generalize the
construction for quantum multiple tori.
\end{abstract}


\section{Introduction}
Noncommutative geometry is a rich area of research, which extends
the classical notions of topology, differential geometry and group
theory to noncommutative objects and their symmetries.

The latter topic, which includes the theory of Hopf algebras, Hopf
groupoids etc., has been extensively studied in the recent years
both in the direction of Lie-type objects as well as finite dimensional
objects.  One of the ambitious tasks, comparable with the classification
of finite groups might be the classification of finite-dimensional
semisimple Hopf star algebras. Only partial results for low dimensions
and special cases are know so far (see \mycite{And,Mon} for a review).

One of best-known examples of semisimple finite Hopf algebras
is the Kac-Paljutkin algebra, especially its lowest dimensional
selfdual example $A_8$ \mycite{Kac}.

In this paper we shall demonstrate that they arise from the exact
sequences of Hopf algebras built using the double noncommutative
torus  \mycite{HaMa}. We shall generalize this construction and
provide other examples.

The paper is organized as follows, first, we briefly review the
construction of the double torus and its dual, then we construct
Kac-Paljutkin algebras through exact sequences (for the
deformations at roots of unity) and in the second part of the
paper we generalize the construction and discuss new examples.

\section{The quantum double torus}

The noncommutative torus, although it has the same symmetries
as the commutative one, is no longer a Hopf algebra.  However,
it appears that an interesting Hopf algebra structure exists on
the direct sum of the deformed and nondeformed torus.

The detailed construction is based on the cocycle deformation of
the tensor product of $C(Z_2) \ot C(T^2)$ \mycite{HaMa}, however,
here, for simplicity, we shall use a straightforward description
with generators.

\begin{definition}
Let $DT_q$ be an unital algebra generated by two unitaries $U,V$
and two selfadjoint projectors $P_+,P_-$, satisfying the relations:
\begin{equation}
\begin{array}{l}
P_+ + P_- = 1, \mylabel{r1} \\
P_+ U = UP_+, \\
P_+ V = VP_+,\\
VU = P_+ UV + q P_- UV,
\end{array}
\end{equation}
where $|q|=1$.
\end{definition}
In fact, the introduction of $P_\pm$ as generators is not necessary,
as we may use the following relations:
\begin{eqnarray}
VU V^{-1}U^{-1} - 1  = (q-1) P_-, \\
 VU V^{-1}U^{-1} - q  = (1-q) P_+.
\end{eqnarray}

\begin{lemma}[see \mycite{HaMa}]
The algebra $DT_q$ is a Hopf algebra with the following coproduct, counit
and antipode:
\begin{equation}
\begin{array}{l}
\cop U = U \ot UP_+ + V \ot UP_- ,\\
\cop V = V \ot VP_+ + U \ot VP_-,\\
\cop P_+ = P_+ \ot P_+ + P_- \ot P_-, \\
\epsilon(V)=1, \mylabel{rel1} \\
\epsilon(U)=1, \\
\epsilon(P_+) =1, \\
 S U = U ^{-1} P_+  + V ^{-1} P_- ,\\
 S V = V^{-1}  P_+ + U^{-1} P_-, \\
 S P_+ = P_+.
\end{array}
\end{equation}
\end{lemma}

Verification of all Hopf algebra properties is left to the reader,
equivalently one may use \mycite{HaMa} and the identification of
the generators $U,V,P_\pm$ with the generators of the tensor
algebra $T \otimes \C(\Z_2)$\footnote{Note  that the parameter
$q$ used here corresponds to $q^2$ of \mycite{HaMa}.}.

\begin{observation}
$DT_q$ is a $*$-Hopf algebra.
\end{observation}
Indeed, another simple exercise verifies that for the generators we have:
$$\cop (a^*) = \coo{a}^* \otimes \cot{a}^*.$$

In fact, using the operator $U_\pm = U P_\pm$ and $V_\pm =V P_\pm$
we may see that the $C^\ast$ algebra generated by $DT_q$ is
a compact matrix quantum group (compact matrix pseudogroup)
in the sense of Woronowicz:
\begin{equation}
\cop \left( \begin{array}{cc} U_+ & V_- \\ U_- & V_+ \end{array} \right)
= \left( \begin{array}{cc} U_+ & V_- \\ U_- & V_+ \end{array} \right)
\otimes \left( \begin{array}{cc} U_+ & V_- \\ U_- & V_+ \end{array} \right).
\end{equation}
We leave the verification of the remaining necessary conditions
(see \mycite{Wor}) to the reader.

The theory of unitary representations has been discussed in
\mycite{HaMa}:
\begin{lemma}
The representation of the algebra $DT_{q}$ on the Hilbert space
$\CH = l_2 (\Z^2) \oplus l_2(\Z^2)$ is given by:
\begin{equation}
\begin{array}{l}
U |n,m, \pm \rangle =  |n+1,m, \pm \rangle,\nonumber \\
V |n,m, + \rangle =  |n,m+1, + \rangle,\nonumber \\
V |n,m, - \rangle  =  q^{n} |n,m+1, - \rangle \mylabel{rep}.
\end{array}
\end{equation}
\end{lemma}

Of course, the operators $P_\pm$ act as projections on $\CH_\pm$.

\section{The dual Hopf algebra of the quantum double torus.}

To construct the dual Hopf algebra we have to choose the
appropriate mathematical setup. Although it is not a problem to
define linear functionals on $DT_q$ and their algebraic structure,
the notion of the coproduct requires caution. It seems to us that
the definition of multiplier Hopf algebras \mycite{AvD} is best
suited for the example. We shall recall the necessary definitions
when necessary.

First, let us introduce the basis of the dual algebra generators
$ c_-^{mn}, c_+^{mn}$ defined as linear functionals:
\begin{equation}
\begin{array}{ll}
c_+^{mn} ( U_+^k V_+^l ) = \delta^{mk} \delta^{nl},  &
c_+^{mn} ( U_-^k V_-^l ) = 0, \\
c_-^{mn} ( U_+^k V_+^l ) = 0, & c_-^{mn} ( U_-^k V_-^l ) =
q^{-\oh mn} \delta^{mk} \delta^{nl}.
\end{array}
\end{equation}

We have chosen here a nontrivial factor in the duality relation, which
corresponds to the rescaling of the generators in order to obtain
simpler algebraic relations. Indeed, we could translate the
multiplication and comultiplication rules to the dual algebra and
obtain:
\begin{equation}
\begin{array}{l}
c_+^{kl} c_+^{mn} = \delta^{km} \delta^{ln} c_+^{mn}, \mylabel{dual} \\
c_+^{kl} c_-^{mn} = \delta^{kn} \delta^{lm} c_-^{mn}, \\
c_-^{kl} c_+^{mn} = \delta^{km} \delta^{ln} c_-^{mn}, \\
c_-^{kl} c_-^{mn} = \delta^{kn} \delta^{lm} c_+^{mn}.
\end{array}
\end{equation}

Then we define the dual algebra $(DT_q)^*$ as an algebra spanned
by linear combinations of finitely many generators $c_\pm^{ij}$.
In fact the algebra (\ref{dual}) might be identified with the
crossed-product algebra of the group $\Z_2$ by the algebra of
functions of compact support on $\Z^2$.

Let us recall a definition of comultiplication for the multiplier
Hopf algebra $A$:
\begin{definition}[see \mycite{AvD}] \label{def31}
A comultiplication is a $\ast$-homomorphism
$\cop: A \to M(A \otimes A)$ so that $\cop(a)(1 \otimes b)$ and
$(a \otimes 1)\cop(b)$ are in $A \otimes A$ and $\cop$ is
coassociative in the sense that for every $a,b,c \in A$
$$ (a \otimes 1 \otimes 1)(\cop \otimes \id)(\cop(b)(1 \otimes c)
= (\id \otimes \cop)((a \otimes 1)\cop(b))(1 \otimes 1 \otimes c).$$
\end{definition}

For the algebra $(DT_q)^*$ we have:

\begin{definition}
The coproduct is as follows:
\begin{equation}
\begin{aligned}
\cop c_+^{mn} &= \sum_{i+k=m} \;\; \sum_{j+l=n} c_+^{ij} \otimes c_+^{kl}, \\
\cop c_-^{mn} &=  \sum_{i+k=m} \;\; \sum_{j+l=n} q^{\oh (jk-il)}
c_-^{ij} \otimes c_-^{kl},
\end{aligned}
\end{equation}
then the antipode and counit are:
\begin{eqnarray}
&& \epsilon(c_\pm^{ij}) = \delta_{i0} \delta_{j0}, \\
&& S(c_+^{ij}) = c_+^{(-i)(-j)}, \\
&& S(c_-^{ij}) = c_-^{(-j)(-i)},
\end{eqnarray}
Finally, the star structure:
\begin{eqnarray}
\left(c_+^{ij}\right)^\ast = c_+^{ij}, \\
\left(c_-^{ij}\right)^\ast = c_-^{ji}.
\end{eqnarray}
\end{definition}

Clearly the target space of the comultiplication is contained within
the multiplier algebra of $(DT_q)^* \otimes (DT_q)^*$, it is also
easy to verify that the other conditions are satisfied, for example,
we shall demonstrate here that $\cop(c_+^{ij}) (1 \otimes c_-^{kl})$
is in $(DT_q)^* \otimes (DT_q)^*$:

\begin{eqnarray*}
&& \cop(c_+^{ij}) (1 \otimes c_-^{kl}) =
\sum_{r+s=i} \sum_{t+w=j} (c_+^{rt} \otimes c_+^{sw}) (1 \otimes c_-^{kl}) = \\
&& \phantom{xxx} = c_+^{(i-l)(j-k)} \otimes c_-^{kl}.
\end{eqnarray*}

In fact, one can easily verify that $(DT_q)^*$ is a {\em discrete
quantum group} in the sense of Van Daele's definition \cite{AvD2},
as it is a direct sum of full matrix algebras with a multiplier
Hopf algebra structure.

Since the algebra is contained in its multiplier we might try to
extend the definition of  the comultiplication to $M((DT_q)^*)$.
Let us define three elements $e_1, e_2$ and $\sigma$ from
$M((DT_q)^*)$:

\begin{eqnarray}
e_1 = \sum_{n,m} n c_+^{nm}, \;\;\;\; e_2= \sum_{n,m} m c_+^{nm}, \mylabel{lie1} \\
\sigma =  \sum_{n,m} c_-^{nm}.
\end{eqnarray}

These elements must be understood as belonging to $M(DT_q)^*$,
i.e. through their multiplication on the elements of $(DT_q)^*$.
Notice that $\sigma^2$ is the identity in $M(DT_q)^*$ and all
elements $c_-^{mn}$ could be expressed as a product of $\sigma$
and $e_+^{mn}$:
\begin{equation}
\begin{array}{ll}
{\sigma} c_+^{mn} = c_-^{mn} , &
c_+^{mn} \sigma =  c_-^{nm}.
\end{array}
\end{equation}

Let us examine the algebra generated by $e_1, e_2$ and $\sigma$
and the extension of the comultiplication on them:
\begin{equation}
\begin{array}{l}
[e_1, e_2] = 0, \\
e_1 \sigma = \sigma e_2, \mylabel{sig} \\
\sigma^2 = 1, \\
\cop e_i = e_i \otimes 1 + 1 \otimes e_i, \\
\cop \sigma = (\sigma \otimes \sigma) (q^{\oh (e_2 \otimes e_1- e_1 \otimes e_2)}).
\end{array}
\end{equation}

Note that the expression $q^{\oh (e_2 \otimes e_1 -e_1 \otimes
e_1)}$ makes sense as an element of the multiplier algebra of the
tensor product, indeed, taking the example $c_+^{ij} \otimes
c_+^{kl}$, for instance, we have:

$$ q^{\oh (e_2 \otimes e_1 -e_1 \otimes e_1)} c_+^{ij} \otimes c_+^{kl} =
q^{\oh (jk-il)} c_+^{ij} \otimes c_+^{kl} \in (DT_q)^* \otimes (DT_q)^*.$$

The elements $e_1,e_2$ are self-adjoint. Out of the above
relations we might immediately obtain:
\begin{lemma}
The algebra defined above is a twist of the cocommutative crossed
product of $u(1) \times u(1)$ with $\C\Z_2$ by the Cartan element
$q^{\oh(e_1 \ot e_2)}$.
\end{lemma}
\begin{proof}
Indeed, using (\myref{sig}) we have:
$$ q^{\oh(e_1 \ot e_2)} (\sigma \ts \sigma) = (\sigma \ts \sigma)
q^{\oh(e_2 \ot e_1)},$$
so:
$$ \cop \sigma = (\sigma \otimes \sigma)
\left(q^{\oh (e_2 \otimes e_1- e_1 \otimes e_2)} \right) =
q^{\oh(e_1 \ot e_2)} (\sigma \ts \sigma)
q^{-\oh(e_1 \ot e_2)}. $$

Since $e_1$ and $e_2$ commute with each other their coproduct
does not change.
\end{proof}
\subsection{The action of the dual on the quantum torus}

Similarly as in the classical case of the Lie group of the torus
we might interpret the dual algebra (and its multiplier) as the
symmetry algebra of the $DT_q$, in terms of the action of the Hopf
algebra on the algebra of the double quantum torus:

\begin{equation}
\begin{array}{ll}
e_1 \acts U_\pm = U_\pm, & e_1 \acts V_\pm = 0, \\
e_2 \acts U_\pm = 0, & e_2 \acts V_\pm = V_\pm, \\
\sigma \acts U_\pm = V_\mp, & \sigma \acts V_\pm = U_\mp.
\end{array}
\end{equation}

\section{Finite extensions of the torus.}

So far, we have been treating $q$ as a generic parameter,
$|q|=1$. An interesting situation occurs, however, when
$q$ is a root of unity.  In fact, when $q=1$, $DT_q$ is
a commutative algebra, which is the algebra of functions
on the twisted double torus:

\begin{observation}
Let $G$ be the crossproduct group $T^2 \rtimes \Z_2$ with the
nontrivial action of $\Z_2$ on $T^2$ by the flip:
$$ \sigma \left( \begin{array}{c} z \\ w \end{array} \right) =
\left( \begin{array}{c} w \\ z \end{array} \right),
$$
if $\sigma$ is the generator of $\Z_2$.

With the multiplication on $G$
$$ \left( \left( \begin{array}{c} z \\ w \end{array} \right), \sigma \right)
\cdot \left( \left( \begin{array}{c} z' \\ w' \end{array} \right), \sigma' \right)
= \left( \left( \begin{array}{c} z \\ w \end{array} \right) \cdot
\sigma \left( \begin{array}{c} z' \\ w' \end{array} \right), \sigma \sigma'\right),$$

we have a group structure on the space, which topologically is a disjoint
sum of two tori. The algebra of continuous functions on $G$ is a Hopf
algebra and it is identified with the $C^*$ algebra $DT_{q=1}$.
\end{observation}

The proof of both group structure and the identification of the
obtained Hopf algebra is an easy exercise and is left to the reader.

If $q$ is a primitive nontrivial root of unity so that $q^N=1$,
$N \geq2$ we have the following:

\begin{lemma}
Let $j$ be a map $j: DT_q \to DT_q$ defined as identity on polynomials
of order less than $N$ in $U$ and $V$, and
$j(U^N)=1$ and $j(V^N)=1$,
and extended as the algebra homomorphism on the
entire $DT_q$.  Then, if $q^N=1$, $j$ is an Hopf
algebra homomorphism.
\end{lemma}

\begin{proof}
First, we check that the definition is correct - indeed, since
$U^N$ and $V^N$ are invertible elements from the center we might
map it to the unit of the algebra.  Next, we verify:
$$
\begin{array}{l}
\cop 1 = 1 \otimes 1 = j(U^N) \otimes
j(U^N) P_+ +  \\
\phantom{xxxxxxx} + j(V^N) \otimes j(U^N)P_- =\cop j(U^N),
\end{array}
$$
and, similarly, one can prove the relation for
$V^N$.
\end{proof}

\begin{corollary}
The image $j(DT_q)$ is an Hopf algebra $T_q^f$.
\end{corollary}

Using the map $j$ we might construct the exact sequences of Hopf
algebras:
\begin{equation}
DT_q^N \mapright{i}  DT_q  \mapright{j} T_q^f,
\end{equation}
where  $DT_q^N$ denotes the subalgebra generated by $U^N$
and $V^N$ (it is easy to demonstrate that it is a sub-Hopf algebra),
$i$ is the inclusion and $j$ the above mentioned surjection.
All maps are, of course, Hopf algebra homomorphisms and the
composition $j \circ i$ gives the counit:
\begin{equation}
j \circ i (a) = 1 \epsilon(a),  1 \in T_q^f, \forall a \in DT_q^N,
\end{equation}

What remains to prove is that the kernel of $j$ is exactly the
Hopf ideal of the form $DT_q i (DT_q^N)^+ DT_q$, where
$(DT_q^N)^+$ is the kernel of the counit map on $DT_q^N$.

The inclusion is obvious by construction, so let us assume that $a
\in DT_q$ is such that $j(a) = 0$. We shall restrict ourselves in
the proof to the case when $a$ is a polynomial in $U,V$, in fact,
it would be easier to use $U_+,U_-, V_+,V_-$, for which the map
$j$ is given as:
$$ j(U^N_+) = P_+, \;\; j(U^N_-) = P_-, \;\;  j(V^N_+) = P_+,
\;\; j(V^N_-) = P_-.$$

Since $P_+$ and $P_-$ are projections commuting with the entire
algebra we consider them and the corresponding subalgebras
separately for a while (the kernel of $j$ splits as well). For
$P_+$ we have a completely commutative situation and the kernel of
$j$ (restricted to $P_+ DT_q$) is characterized by the condition
that a polynomial of two variables $x,y$ vanishes for every pair
of $N$-th roots of unity $x,y \in \C, x^N=y^N=1$. Clearly, it must
be of the form:
$$ (x^N-1) p(x,y) + (y^N-1)r(x,y), $$
and thus it is an element of the ideal generated by $i(U^N) -1$
and $i(V^N)-1$.

For the $P_-$ part we have a noncommutative situation. Take a
polynomial $p(U_-,V_-)$ and assume that $j(p)=0$. If
$p = \sum a_{nm} U_-^n V_-^m$ then it means that for every
$0 \leq i,j < N$ we have:
$$ \sum_{r,s} a_{(i+rN)(j+sN)} = 0. $$
Let us fix $i,j$. Since only a finite number of $a_{(i+rN)(j+sN)}$
is different from zero the corresponding part of the polynomial
$p(U_-, V_-)$ could be written as:
$$ \left( \sum_{r,s} a_{(i+rN)(j+sN)} U_-^{rN} V_-^{sN} \right) U_-^i V_-^j, $$
where we have used the fact that $q^N=1$ and therefore $U_-^N$ and
$V_-^N$ commute with all elements. Note that we have again reduced
our problem to the commutative one: we have a polynomial in
$U_-^N, V_-^N$ such that the sum of its all coefficients vanishes.
We leave to the reader the verification that it could be written
as:
$$ ( U_-^N-1 )  p_1(U_-^N, V_-^N) +  ( V_-^N-1 )  p_2(U_-^N, V_-^N), $$
for some polynomials $p_1,p_2$, and this clearly belongs to the
ideal generated by $i(U^N)-1$ and $i(V^N)-1$.

Since we can view the algebra $DT_q^N$ as an algebra of
functions on the torus (commutative algebra generated by two
unitaries), the finite Hopf algebra $T_q^f$ could be interpreted
as functions on the noncommutative "fibre", which extends the
classical torus to the quantum double torus at roots of unity.

\begin{lemma}
The algebra $T_q^f$ is a finite dimensional semisimple $C^*$ algebra,
and as an algebra over $\C$ is isomorphic to $M_N(\C) \oplus \C^{N^2}$.
\end{lemma}

\begin{proof}
The algebra is described through the relations (\myref{r1}) with
the additional constraints $U^N=V^N=1$. Clearly, the projections
$P_+$ and $P_-$ define two ideals, a commutative and a
noncommutative one. When we restrict ourselves to the algebra of
the noncommutative ideal, then $P_-$ becomes a unit and the
algebra of this ideal is  generated by two $q$-commuting, $q^N=1$,
unitaries, hence it is a complex matrix algebra of dimension $N$.
\end{proof}

\begin{lemma}[see \mycite{Sek},\mycite{Vai}]
The algebra $T_q^f$ for $q^N=1$ is isomorphic to the
Kac-Paljutkin finite Hopf algebra of rank $2N^2$.
\end{lemma}

For the case of the dual Hopf algebra of the quantum double torus
we find as well an embedded Hopf subalgebra (since we work
with infinite sums the result makes sense only as a multiplier
algebra):

\begin{lemma}
Let $w_\pm^{ij}$, $ 0 \leq i,j < N$, be the elements
defined as formal series:
$$ w_\pm^{ij} = \sum_{r,s} c_\pm^{(i+Nr),(j+sN)},$$
Then the subalgebra ${\mathcal W}$ generated by
$w_\pm^{ij}$ is a Hopf subalgebra.
\end{lemma}
\begin{proof}
First, let us notice that the commutation
relations will be identical as for $c_\pm^{ij}$:
\begin{equation}
\begin{array}{l}
w_+^{kl} w_+^{mn} = \delta^{km} \delta^{ln} w_+^{mn} \\
w_+^{kl} w_-^{mn} = \delta^{kn} \delta^{lm} w_-^{mn} \\
w_-^{kl} w_+^{mn} = \delta^{km} \delta^{ln} w_-^{mn} \\
w_-^{kl} w_-^{mn} = \delta^{kn} \delta^{lm} w_+^{mn}, \\
 \left(w_+^{ij}\right)^\ast = w_+^{ij} \\
\left(w_-^{ij}\right)^\ast =  w_-^{ji}.
\end{array}
\end{equation}

It only remains to prove whether the coproduct closes
within \mbox{${\mathcal W} \otimes {\mathcal W}$}(we use
the formal expressions within the multiplier of the tensor
product):
\begin{eqnarray*}
\cop w_+^{mn} &=& \sum_{r,s} \cop c_+^{(m+Nr),(n+sN)} \\
&=& \sum_{r,s} \sum_{i+k=m+Nr} \sum_{j+l=n+sN} c_+^{ij} \otimes c_+^{kl} \\
&=& \sum_{r,s} \sum_{i,j}  c_+^{ij} \otimes c_+^{(m+Nr-i)(n+sN-j)} \\
&=& \sum_{i,j}  c_+^{ij} \otimes w_+^{([m-i])([n-j])} \\
&=& \sum_{0 \leq i,j < N} w_+^{[i][j]} \otimes w_+^{([m-i])([n-j])} \\
&=& \sum_{[i'+k']=m} \sum_{[j'+l']=n}  w_+^{i'j'} \otimes
w_+^{k'l'},
\end{eqnarray*}
where we use $[x]$ to denote the operation modulo $N$ and the last
sum is over indices $i',j',l',k'$, which run from $0$ to $N-1$.
Of course, a similar calculation could be done for $w_-^{ij}$.
\end{proof}

The algebra has $2N^2$ generators and the following proposition
shows its structure:
\begin{lemma}
For fixed $i \not=j$, the linear span of generators $w_\pm^{ij}, w_\pm^{ji}$
forms a maximal ideal within ${\mathcal W}$, and the $\ast$-representation
of these generators on $\C^2$ is:
\begin{eqnarray*}
w_+^{ij} = \left( \begin{array}{cc} 1 & 0 \\ 0 & 0 \end{array} \right), &
w_+^{ji} = \left( \begin{array}{cc} 0 & 0 \\ 0 & 1 \end{array} \right), \\
w_-^{ij} = \left( \begin{array}{cc} 0 & 0 \\ 1 & 0 \end{array} \right), &
w_-^{ji} = \left( \begin{array}{cc} 0 & 1 \\ 0 & 0 \end{array} \right).
\end{eqnarray*}
\end{lemma}

\begin{lemma}
For $i=j$, the linear span of $w_\pm^{ii}$ generates a commutative
subalgebra, with the following representation:
\begin{eqnarray*}
w_+^{ii} = \left( \begin{array}{cc} 1 & 0 \\ 0 & 1 \end{array} \right), &
w_-^{ii} = \left( \begin{array}{cc} 1 & 0 \\ 0 & -1\end{array} \right).
\end{eqnarray*}
\end{lemma}
The proof of both lemmas is straightforward
and therefore we skip it.

\begin{observation}
As an algebra, ${\mathcal W}$ is isomorphic to:
$$  M_2(\C) \oplus \cdots \oplus M_2(\C) \oplus \C \cdots \C,$$
with $\frac{1}{2} N(N-1)$ copies of $M_2(\C) $ and $2N$ copies
of $\C$.
\end{observation}

\begin{corollary}[see again \mycite{Sek,Vai}]
The finite-dimensional Hopf algebra ${\mathcal W}$ is the dual to
the Kac-Paljutkin algebra.
\end{corollary}
\section{The multidimensional generalization}

Let $G$ be a finite group, which we treat as a subgroup of a
permutation group, so for $h \in G$ and $i \in \Z^n$\footnote{Note that
in this section $i,j,k,\ldots$ will be multiindices unless stated
otherwise, and not integers as in previous section.} $h(i) $ will denote
the permutation of the multiindex $i$. Let $C^{i}_g$, $g \in G$ $i
\in \Z^n$ be the generators of the associative algebra $\CT_G$:
\begin{equation}
C^{i}_g C^j_h = \delta^{i,h(j)} C^j_{(gh)}, \mylabel{rel2a}
\end{equation}
Now, we can introduce a nontrivial coproduct
structure on the algebra (again, valued in the multiplier of the
tensor product of $\CT_G$) :
\begin{equation}
\cop C^k_g = \sum_{i+j=k} \alpha^{ij}_g C^i_g \otimes C^j_g, \mylabel{rel2}
\end{equation}
where the complex valued coefficients $\alpha^{ij}_g$ must
satisfy:
\begin{eqnarray}
&& \alpha^{ij}_g \alpha^{(i+j)k}_g =\alpha^{i(j+k)}_g \alpha^{jk}_g \mylabel{coa} \\
&& \alpha^{ij}_{gh} = \alpha^{ij}_h \alpha^{h(i) h(j)}_{g}. \mylabel{hom}
\end{eqnarray}
Then (\myref{coa}) guarantees that the coproduct is coassociative
(in the sense of definition \ref{def31}) and (\myref{hom}) that it
is an algebra homomorphism.

The first condition can be solved immediately using
a bilinear form on the $\Z^n$. If we assume:
\begin{equation}
\alpha^{kl}_g = e^{2\pi i \theta_g(k,l)},
\end{equation}
where $\theta_g$ is a bilinear form then (\myref{coa}) is satisfied
automatically and (\myref{hom}) leads to:
\begin{equation}
\theta_{gh}(m,n) = \theta_h(m,n) + \theta_g(h(m), h(n)), \mylabel{cond1}
\end{equation}
for all $g,h \in G$ and $m,n \in \Z^n$. Clearly, for the neutral
element $e \in G$ we have $\theta_e \equiv 0$.

Note that the condition (\myref{cond1}) is a cocycle condition valued
in the bilinear forms. For simplicity we shall denote the latter space
by $\CB$. Consider for a given $k$ all maps from $G^k$ to $\CB$,
this forms a linear space $\CC^k$ and let us define
a linear map $\delta: \CC^k \to \CC^{k+1}:$
\begin{equation}
\begin{aligned}
\delta \phi(g_1, \ldots, g_{k+1}) =& \phi(g_2,\ldots g_{k+1}) + \\
&  \sum_{i=1}^{k} (-1)^i \phi(g_1,\ldots, g_i g_{i+1}, \ldots, g_{k+1}) + \\
& (-1)^{k+1} g_{k+1} \triangleright \phi(g_1, \ldots, g_k).
\end{aligned}
\end{equation}
where $g \triangleright \phi$ denotes the left action of $G$ on $\CB$.
This defines a group cohomology (as $\delta^2 \equiv 0$) with
values in the group module $\CB$ with the trivial right module
multiplication  and the left multiplication (action) defined by the
representation (permutation) of the group $G$.

Thus our condition (\myref{cond1}) is nothing but a cocycle condition,
for $\theta$ understood as a map $G \to \CB$:
$$ \delta \theta = 0, $$
so we can immediately provide at least trivial solutions
$\theta = \delta \phi$, where $\phi$ is an element of $\CB$.

To restrict the freedom depending on the choice of the basis
we observe:
\begin{lemma}
The map:
$$ C^j_g \mapsto e^{- \pi i \theta_g(j,j)} C^j_g = \widetilde{C^j_g},$$
is an algebra automorphisms, which changes the coalgebra
structure so that:
\begin{equation}
\cop \widetilde{C}^k_g = \sum_{i+j=k} \widetilde{\alpha}^{ij}_g
\widetilde{C}^i_g \otimes \widetilde{C}^j_g, \mylabel{rel3}
\end{equation}
where
$$ \widetilde{\alpha}^{ij}_g = e^{\pi i (\theta_g(i,j) - \theta_g(j,i))},$$
so $\widetilde{\theta_g^{ij}}$ is then antisymmetric for every $g$.
\end{lemma}

Hence, without loosing any information, we can always restrict
ourselves to antisymmetric cocycles $\theta_g$ and from now on we
shall use the notation without tilde keeping in mind that this is
already a basis in which $\theta$-s are antisymmetric.

Further, having produced a coproduct we shall
demonstrate the existence of the counit and the antipode.
\begin{lemma}
With the above coproduct, counit:
\begin{equation}
\epsilon(C^i_g) = \delta^{i,0},
\end{equation}
and the antipode:
\begin{equation}
S C^k_g = C^{-g(k)}_{g^{-1}},
\end{equation}
as well as with the star structure:
\begin{equation}
( C^i_g)^\ast = C^{g(i)}_{g^{-1}}, \mylabel{ast}
\end{equation}
it is a $\ast$-Hopf algebra provided that
\begin{equation}
\theta_g(i,j)^\ast = - \theta_{g^{-1}}(g(i),g(j)),
\end{equation}
however, one can easily notice that this is guaranteed by (\myref{cond1})
if the forms $\theta$ are real.

The algebra $\CT_G$ is a $\ast$-Hopf algebra (in the sense of
multiplier Hopf algebras) and - similarly as for the dual of the
double torus - it is an example of a discrete quantum group
\mycite{AvD2}.

\end{lemma}

\begin{observation}\mylabel{obs3}
For $\Z_N$ the structure of the coproduct is set by
a single bilinear antisymmetric form (matrix)
$\Theta=\theta_1$. Its matrix elements have to satisfy
(addition is mod $N$):
\begin{equation}
\sum_{k=0}^{N-1} \Theta_{i+k,j+k} = 0.
\end{equation}
for every $0 \leq i,j < N$. This follows directly from
(\myref{cond1}).
\end{observation}

\begin{example}
For the group $\Z_2$ the relations for the matrix $\Theta=\theta_1$ are:
\begin{equation}
\begin{array}{l}
\Theta_{12}+\Theta_{21}=0, \\
\Theta_{11}+\Theta_{22}=0.
\end{array}
\end{equation}
and they are automatically satisfied for any antisymmetric matrix:
\begin{equation}
\Theta = \left( \begin{array}{cc} 0 & \theta \\ -\theta & 0
\end{array} \right), \mylabel{z2sol}
\end{equation}
The solution (\myref{z2sol}) is a trivial cocycle, it is easy
to see that it is $\delta$ of a constant bilinear form and the algebra
is the dual of the double torus as described by (\myref{dual}).
\end{example}

\begin{example}
Let us consider the simplest nontrivial example -- apart from
the known $\Z_2$ case - for the group $\Z_3$. As we have
observed in (\myref{obs3}) we are left with the following
restriction for the matrix elements of $\Theta$ (we have
already used its antisymmetry):
\begin{equation}
\Theta_{01}+\Theta_{12} + \Theta_{20}=0,
\end{equation}
and the solution is given by:
\begin{equation}
\Theta = \left(
\begin{array}{ccc}
0 & \theta & -\rho \\ -\theta & 0 & -(\theta+\rho) \\ \rho & (\theta+\rho) & 0
\end{array} \right)
\end{equation} \mylabel{eq38}
\end{example}

\subsection{The Lie algebra picture}
Let us define, similarly as for the double torus case
(\myref{lie1}) a formal construction of the convenient
generators within the multiplier algebra:
\begin{equation}
e_j = \sum_{i \in \Z^n} i_j C^i_e, \;\; j =1,2,\ldots,n.
\end{equation}
and the elements $\tilde{g}$, labelled by the elements of the
group $G$:
\begin{equation}
\tilde{g} = \sum_{i \in \Z^n}  C^i_g.
\end{equation}

Using the relations (\myref{rel2a}) we verify the algebra
structure:
\begin{equation}
\begin{array}{l}
[ e_i, e_j ] = 0 \\
\tilde{g}\tilde{h} = \widetilde{(gh)} \\
\tilde{g} e_i = e_{g(i)} \tilde{g}
\end{array}
\end{equation}

However, the structure of the coproduct is more complicated:
\begin{equation}
\begin{array}{l}
\cop e_i = e_i \ts 1 + 1 \ts e_i \\
\cop \tilde{g} = e^{2\pi i \Theta_{kl} e_k \ts e_l} \tilde{g} \ts
\tilde{g}, \mylabel{cop2}
\end{array}
\end{equation}
where $\Theta$ is the matrix of the antisymmetric form $\theta$.
The formula (\myref{cop2}) can arise from a twist only if
$\theta = \delta \phi$, so $\theta$ is a trivial cocycle. Then:
\begin{equation}
\cop \tilde{g} = e^{2\pi i \phi_{kl} e_k \ts e_l} \tilde{g} \ts \tilde{g}
e^{-2\pi i \phi_{kl}  e_k \ts e_l},
\end{equation}

We can summarize that for the trivial cocycle the presented
deformation is a twist of the cross-product of multiple
dual tori Hopf algebras by a finite group.

\subsection{The dual algebra}
Similarly as in the original $\Z_2$-case we might construct
the dual Hopf algebra. We introduce the algebra elements
$U^i_g$, $g \in G$, $i \in \Z^n$ and by duality:
$$ \langle C^i_g, U^j_h \rangle = \delta_{gh} \delta^{ij},$$
we obtain from (\myref{rel2a}) and (\myref{rel2}) both the algebra
and the coalgebra structure. Again, we restrict ourselves only to
the case of the antisymmetric matrix $\theta_g$:
\begin{equation}
\begin{aligned}
U^i_g U^j_h = \delta_{gh} e^{2\pi i \theta_g(i,j)} U^{i+j}_g, \\
\cop U^i_f = \sum_{gh=f}  U^{h(i)}_g \ts U^i_h.  \mylabel{rel-u1}
\end{aligned}
\end{equation}

The counit, antipode and the star structure are, respectively,
given by:
\begin{equation}
\begin{array}{l}
\epsilon( U^j_g) = \delta_{ge}, \\
S (U^j_g) = U^{-g(j)}_{g^{-1}},\\
(U^j_g )^\ast = U^{-j}_g. \mylabel{rel-u2}
\end{array}
\end{equation}

where $e \in G$ is the neutral element of the group and we have,
of course,  identified $\sum_{g \in G} U^0_g$ with the unity in the
algebra. Of course, each $U^0_g$ is a central, self-adjoint projector
in the algebra.

\begin{observation} The algebra spanned by
$U_g^i$ could be also viewed as generated by
unitaries. If $v_1,\ldots, v_n$ are the basis
of $\Z^n$ then:
$$ U^{v_i} = \sum_{g \in G} U^{v_i}_g,$$
are unitary elements of the algebra, and:
$$ (U^{v_m})^* (U^{v_n})^* U^{v_m} U^{v_n} =
\sum_g e^{4\pi i \theta_g(v_m,v_n)} U_g^0. $$
\end{observation}
\section{Finite subalgebras}

Similarly as in the $\Z_2$ case, where for $q$ being a simple root
of unity we have constructed a finite dimensional Hopf
subalgebra, we might attempt to repeat the procedure here.

Let us assume that the matrices $\theta$ are in $M_n(\Q)$ and $L$
be a lattice in $\Z^n$, such that for every $g \in G$ and every $i
\in L$ and arbitrary $j \in \Z^n$ we have:
$$ \theta_g(i,j) \in \Z.$$

Clearly, $L$ must be a subring of $\Z^n$, which
is invariant by the action of $G$. Then we have:
\begin{lemma}
Let ${\mathcal I}$ be the quotient  ${\mathcal I}=\Z^n/L$.
Then we might define a Hopf algebra by the following relations:
$$ w^{[i]}_g = \sum_{p \in L} C^{i+p}_g, \;\; [i] \in {\mathcal I}.$$
\end{lemma}
It is an easy exercise to prove that this is a sub algebra of $\CT_G$:
\begin{equation}
\begin{array}{l}
w^{[i]}_g w^{[j]}_h = \sum_{p, s \in L} C^{i+p}_g C^{j+s}_h =  \\
\phantom{xx} = \sum_{p, s \in L} \delta_{(i+p), h(j+s)} C^{j+s}_{(gh)}
= \sum_{s \in L} \delta_{[i], [h(j) + h(s)] } C^{j+s}_{(gh)} = \\
\phantom{xx} = \delta_{[i], h([j])} w^{[j]}_{(gh)}.
\end{array}
\end{equation}

For the coaction we have:
$$\cop w^{[i]}_g = \sum_{s \in L} \sum_{k+l = i +s} e^{2\pi i \theta_g(k,l)} C^k_g \ts C^l_g = \ldots $$
now, since we know that $\theta_g(k,l)$ is integer for every $k \in L$ and arbitrary $l$ the factor
in front of the tensor product depends only on the class of $k$ and [l] in the quotient,
$[k], [l]  \in {\mathcal I}$.
Therefore:
$$ \ldots = \sum_{s \in L} \sum_{t \in L} \sum_{k \in {\mathcal I}} e^{2\pi i \theta_g([k],[i+s-k-t])}
C^{k+t}_g \ts C^{i+s-k-t}_g = \ldots $$
summing up over $s$ and $t$ and keeping in mind that $[i+s-k-t] = [i]-[k]$ we finally obtain:
$$ \ldots = \sum_{[k]+[l]=[i]} e^{2\pi i \theta_g([k],[l])}
C^{[k]}_g \ts C^{[l]}_g, $$
where the sum is over $[k],[l] \in  {\mathcal I}$.

As for the dual case, the arguments follow the line of construction
in Proposition 4.2. We notice that the elements $U^{v_i}$, for $v_i \in L$
form a commutative subalgebra and, since $L$ is invariant with
respect to the action of the group $G$, it is a Hopf-subalgebra
$\CT^L_G \subset \CT_G$.

Then we can construct (similarly as in the Proposition 4.2) an exact
sequence of Hopf algebras:
\begin{equation}
\CT^L_G \mapright{i} \CT_G \mapright{j} \CF^L_G, \mylabel{exseq}
\end{equation}
where $\CF^L_G$ is a finite Hopf algebra, which could be constructed
as an image of the map $\tilde{j}: \CT_G \to \CT_G$ ($j$ is $\tilde{j}$
with target space restricted to the image $\tilde{j}(\CT_G)$):
$$\tilde{j}(U^i_g) = U^{[i]}_g, $$
where $[i] \in \Z^n/L$, so in particular for every $k \in L$:
$$\tilde{j}(U^k_g) = U^{0}_g, $$
and
$$\tilde{j}(U^k) = 1,$$
where $U^k$ denotes $\sum_g U^k_g$.

It is easy to verify that $ \tilde{j}$ is an Hopf algebra
homomorphism, so its image is a Hopf algebra. It only
remains an easy exercise to verify that the sequence
(\myref{exseq}) is exact.

\subsection{The example of $\Z_3$.}

We shall present here the application of the results
obtained earlier for the case of the group $G=Z_3$
acting by cyclic permutation.

As we have already demonstrated in (38) the structure of
the Hopf algebra is set by an antisymmetric matrix with two
arbitrary parameters $\theta$ and $\rho$ (\myref{eq38}).

We shall assume now that both $N\theta$ and $N\rho$ are integer
for some $N \in \Z$ (we take, of course the smallest possible
$N$). Define the lattice $L$ as $N\Z^3$, i.e., all integer
multiindices, such that every component is divisible by $N$.

Let us denote the generators of our finite Hopf-algebra
by $U_0^v$, $U_1^v$ and $U_2^v$, the indices corresponding to
the elements of the group $\Z_3$ and $v$ taking values in the three
standard basis vectors of $\Z^3$, for simplicity we shall abbreviate
the notation and denote $U_i^{e_k}$ as $U_i^k$. We shall also use
the projectors on each of the component denoted respectively
$P_i$.

Clearly, according to (\myref{rel-u1})  $U_j^v  U_k^w = 0$ if $j
\not=k$, so we shall have a direct sum of three subalgebras,

The commutation relations within each of them are as follows:
\begin{itemize}
\item $g=0$
$$ U^i_0 U^j_0 = U^j_0 U^i_0,  \;\;\;\; (U_0^i)^N = P_0. $$

\item $g=1$
$$
\begin{aligned}
U_1^1 U_1^2 =  q^a  U_1^2 U_1^1, \\
U_1^1 U_1^3 =  q^b  U_1^3  U_1^1, \\
U_1^2 U_1^3 =  q^{b-a}  U_1^3  U_1^2,\\
(U_1^j)^N = P_1,
\end{aligned}
$$
where $q^N=1$, $a = N \theta$, $b = - N \rho$.

\item $g=2$
$$
\begin{aligned}
U_2^1 U_2^2 =  q^{b}  U_2^2 U_2^1, \\
U_2^1 U_2^3 =  q^{b-a}  U_2^3  U_2^1, \\
U_2^2 U_2^3 =  q^{-a}  U_2^3  U_2^2, \\
(U_2^j)^N = P_2.
\end{aligned}
$$
\end{itemize}

The first component of the algebra is commutative and is simply
isomorphic to $\C^{(N^3)}$. As for the second and the third let us
notice that the relations strongly depend on $N$ as well as $a$
and $b$.

To provide an explicit simplest but new example, which extends the
results of the double torus and the Kac-Paljutkin algebras, we
shall focus our attention on the cases $N=2$ and $N=3$.

\subsubsection{$N=2$ case}
In this case $q=-1$ and $a,b$ are either $0$ or $1$. Note that the second
and the third component of the algebra ($g=2$) are noncommutative
(apart from the $a=b=0$ trivial case) and, for each of the possible
situations ($a=1,b=0$; $a=0,b=1$; $a=b=1$) we obtain the algebra
$M_2(\C)$ for. Therefore the algebraic structure of the full algebra
is $M_2(\C)^4 \oplus \C^8$.

\subsubsection{$N=3$ case}
In this situation, $q^3=1$ and we have, in principle, 9
possibilities for the $g=1$  (as well as $g=2$) components of the
algebra. Let us analyze the center of the subalgebra in question.
Taking a monomial $(U_1^1)^\alpha (U_1^2)^\beta  (U_1^3)^\gamma$,
we verify that it is in the center if:
\begin{eqnarray}
&& \beta a + \gamma b = 0\;\; (\hbox{mod\ } 3), \\
&& -\alpha a  + \gamma (b-a) = 0 \;\;(\hbox{mod\ } 3), \\
&& - \alpha b - \beta (b-a) = 0 \;\;(\hbox{mod\ } 3).
\end{eqnarray}

Out of this system of linear equations we immediately get that
either $a=b=0$ or $\alpha + \beta + \gamma$ must be $0$ modulo
$3$. Since we are interested only in the nontrivial case we assume
the latter identity. In addition, we have to take one of the
equations, for instance $a \beta + b \gamma = 0 \hbox{mod\ } 3$.

It appears that independently of the values of $a$ and $b$ (apart
from the $a=b=0$ trivial case) the center of the subalgebra is
isomorphic to $\C^3$ and, furthermore, the structure of the
algebra is $M_3(\C) \oplus M_3(\C) \oplus M_3(\C)$.

Since for the third component $g=2$ we have almost the same
system of equations, with the only exception that $a,b$ are
replaced by $b,b-a$ (modulo $3$). Therefore the same result
applies in this case and finally we could summarize that
for the $N=3$ finite fibration of the "triple tori"  we obtain
a Hopf algebra structure for the algebra $\C^{27} \oplus M_3(\C)^6$.

\section{Final remarks}
We have presented in this paper the method of construction
of finite semisimple Hopf algebras through the finite fibrations
of the double torus, then extending the results to the
generalizations - "quantum multiple tori".

We have related the construction with the cohomology of finite
groups, showing that the Hopf algebra structure is a twist of the
crossproduct of the group by the dual tori only if the cocycle
defining the deformation is trivial. It would be an interesting
task to provide an explicit example of such deformation and
study its finite fibrations.

\end{document}